\documentclass[prb,twocolumn,showpacs,preprintnumbers,amsmath,amssymb,superscriptaddress]{revtex4}

\usepackage{graphicx}
\usepackage{dcolumn}
\usepackage{bm}
\usepackage{longtable}

\begin{document}

\preprint{APS/123-QED}

\title{Lattice-form dependent orbital shape and charge disproportionation in charge- and orbital-ordered 
manganites}

\author{D. Okuyama}%
\affiliation{Cross-Correlated Materials Research Group (CMRG), ASI, RIKEN, Wako 
351-0198, Japan}
\affiliation{Multiferroics Project, ERATO, Japan Science and Technology Agency (JST), 
c/o RIKEN, Wako 351-0198, Japan}
\author{Y. Tokunaga}%
\affiliation{Multiferroics Project, ERATO, Japan Science and Technology Agency (JST), 
c/o RIKEN, Wako 351-0198, Japan}
\author{R. Kumai}%
\affiliation{National Institute of Advanced Industrial Science and Technology (AIST), Tsukuba 305-8562, Japan}
\author{Y. Taguchi}
\affiliation{Cross-Correlated Materials Research Group (CMRG), ASI, RIKEN, Wako 
351-0198, Japan}
\author{T. Arima}%
\affiliation{Institute of Multidisciplinary Research for Advanced Materials, Tohoku University, Sendai 980-8577,
Japan}
\affiliation{RIKEN SPring-8 Center, Hyogo, Japan}
\author{Y. Tokura}%
\affiliation{Cross-Correlated Materials Research Group (CMRG), ASI, RIKEN, Wako 
351-0198, Japan}
\affiliation{Multiferroics Project, ERATO, Japan Science and Technology Agency (JST), 
c/o RIKEN, Wako 351-0198, Japan}
\affiliation{National Institute of Advanced Industrial Science and Technology (AIST), Tsukuba 305-8562, Japan}
\affiliation{Department of Applied Physics, University of Tokyo, Tokyo 113-8656, Japan}

\date{\today}

\begin{abstract}
The orbital shapes and charge disproportionations at nominal Mn$^{3+}$ and Mn$^{4+}$ sites 
for the charge- and orbital-ordered phases have been studied on half-doped manganites 
Pr(Sr$_{0.1}$Ca$_{0.9}$)$_2$Mn$_2$O$_7$ and Eu$_{0.5}$Ca$_{1.5}$MnO$_4$ 
with double-layer and single-layer Mn-O networks, respectively, by means of x-ray structural analyses, 
in comparison with Pr$_{0.5}$Ca$_{0.5}$MnO$_3$ with the pseudo cubic network. 
In a single-layer Eu$_{0.5}$Ca$_{1.5}$MnO$_4$ system, the ($y^2-z^2$)/($z^2-x^2$)-type orbital shape is observed, 
while the ($3y^2-r^2$)/($3x^2-r^2$)-type orbital shape in a pseudo cubic Pr$_{0.5}$Ca$_{0.5}$MnO$_3$ system. 
In a double-layer Pr(Sr$_{0.1}$Ca$_{0.9}$)$_2$Mn$_2$O$_7$ system, 
the orbital shape is found to undergo a large change upon thermally induced rotation of orbital stripe. 
Furthermore, clear charge disproportionation is observed for the pseudo cubic and double-layer systems, 
while not in the single-layer system. 
These results indicate that the orbital shape and charge disproportionation are sensitive to the dimension of Mn-O network. 
\end{abstract}

\maketitle


\section{Introduction}

Charge and orbital order (CO-OO) in mixed-valence manganites with perovskite-related structures has been attracting 
great interest since the CO-OO makes a strong impact on crystallography, magnetism, and electrical conduction\cite{Tokura2000}. 
For example, magnetic-field induced melting of long-range and/or short-range CO-OO results in colossal 
magnetoresistance\cite{Tokura2006}. 
Huge changes in electrical conduction due to the melting of CO-OO have been 
reported with the application of other external stimuli, such as an electric field, x-ray, light, and pressure
\cite{Asamitsu1997,Kiryukhin1997,Fiebig1998,Moritomo1997}. 
Many diffraction and theoretical studies have reported the checkerboard type charge order of Mn$^{3+}$ and 
Mn$^{4+}$ ions with charge disproportionation and stripe-type orbital order, 
explaining a complex antiferromagnetic spin order at low temperatures in half-doped manganites
\cite{Wollan1955,Goodenough1955,Chen1996,Murakami1998,Solovyev1999,Mutou1999,Chen1999,Khomskii2000,
Dagotto2001,Popovic2002}. 
On the other hand, another model with less distinct charge disproportionation has also been proposed\cite{Brink1999,Mahadevan2001,Subias1997,Garcia2001,Martin2004,Goff2004,Loudon2005,Milward2005,Cox2008}. 
For example, taking into account the on-site Coulomb interaction, Brink \textit{et al.} have suggested\cite{Brink1999} 
that maximum value of charge disproportionation is as little as 20\%. 
First principle calculation by Mahadevan \textit{et al.} have also shown\cite{Mahadevan2001} that the charge disproprtionation is 
almost negligible for  La$_{0.5}$Sr$_{1.5}$MnO$_4$ compound. 
Herrero-Martin \textit{et al.} have inferred\cite{Martin2004} from the resonant x-ray scattering data 
that the charge disproportionation of Nd$_{0.5}$Sr$_{0.5}$MnO$_3$ is about 20 \%. 
These different models have raised an important question about the nature of CO state. 

Another issue to be clarified  is the orbital shape at Mn$^{3+}$ ion in the CO-OO phase. 
Radaelli \textit{et al.} suggested by powder neutron and synchrotron x-ray diffraction studies 
that the ($3y^2-r^2$)/($3x^2-r^2$)-type 
orbital order takes place in La$_{0.5}$Ca$_{0.5}$MnO$_3$\cite{Radaelli1997}. 
La$_{0.5}$Sr$_{1.5}$MnO$_4$ was also studied as another CO-OO system 
by resonant x-ray scattering and x-ray linear dichroism methods\cite{Murakami1998,Huang2004}; 
the latter \cite{Huang2004} strongly suggested the ($y^2-z^2$)/($z^2-x^2$)-type orbital shape at Mn$^{3+}$ ion. 
It is an unsolved problem why the orbital shape of La$_{0.5}$Ca$_{0.5}$MnO$_3$ and La$_{0.5}$Sr$_{1.5}$MnO$_4$ 
appear different. 
One obvious difference between these two materials is the dimensionality of MnO$_6$ network. 
The variation of the dimensionality has already been reported
\cite{Zimmermann1999,Wakabayashi2001,Zachar2003} to affect the CO-OO state significantly in terms of 
correlation length, but its effect on the orbital shape should be investigated systematically. 
To clarify these issues, we  have made systematic investigations 
on the charge disproportionation and orbital shape in the CO-OO phases of
half-doped manganites with various Mn-O networks (single-, double-, and infinite-layered MnO$_2$ sheets)
by means of x-ray structure analysis. The results indicate that the charge disproportionation is in reality much smaller than unity, 
and that the orbital shape critically depends on the lattice form, 
in particular on the dimensionality of Mn-O network. 
As far as we know, there has been no experimental investigation on the charge disproportionation for the
layered manganites thus far.

The investigated materials in this study are Pr$_{0.5}$Ca$_{0.5}$MnO$_3$, 
Pr(Sr$_{0.1}$Ca$_{0.9}$)$_2$Mn$_2$O$_7$, 
and Eu$_{0.5}$Ca$_{1.5}$MnO$_4$ with pseudo cubic, double-layer, and single-layer Mn-O networks, respectively. 
We chose these materials with good size matching of the ionic radii at the $A$-sites 
to reduce the effect of  quenched disorder (randomness) as much as possible\cite{Tomioka2004,Mathieu2006,Tokunaga2008a}.
Another important point in selecting the target materials  is that the well-defined orthorhombic distortion of all these materials 
enables us to obtain the single-domain orbital-ordered state as locked by the orthorhombicity. 
The concomitant CO-OO is observed at $T_{\mathrm{CO}}$ which is higher than the CE-type antiferromagnetic 
ordering in every case\cite{Tokunaga2008a,Tokunaga2008b}. 
In Pr$_{0.5}$Ca$_{0.5}$MnO$_3$ and Eu$_{0.5}$Ca$_{1.5}$MnO$_4$, the CO-OO transitions take place at 
$T_{\mathrm{CO}}\sim$230 K and 325 K, and the CE-type antiferromagnetic order with spins pointing along the 
$b$-axis is established at $T_{\mathrm{N}}\sim$170 K and 120 K, 
respectively\cite{Jirak1985,Tomioka1996,Jirak2000,Tokunaga2008b}. 
The ferromagnetic zig-zag chain is parallel to the $b$-axis in Pr$_{0.5}$Ca$_{0.5}$MnO$_3$, 
and to $a$-axis  in Eu$_{0.5}$Ca$_{1.5}$MnO$_4$, respectively. 
In contrast, successive CO-OO transitions are observed at $T_{\mathrm{CO1}}\sim$370 K 
and $T_{\mathrm{CO2}}\sim$315 K in Pr(Sr$_{0.1}$Ca$_{0.9}$)$_2$Mn$_2$O$_7$. 
The propagation direction of stripe-type orbital order spontaneously rotates from along the $a$-axis to along the $b$-axis 
at $T_{\mathrm{CO2}}$; this appears as a generic feature of CO-OO double-layer manganites\cite{Tokunaga2008a}. 
CE-type antiferromagnetic order with spins pointing along $b$-axis grows 
below $T_{\mathrm{N}}\sim$153 K\cite{Tokunaga2006}, and the ferromagnetic zig-zag  chain is along $a$-axis. 
In both of the  Eu$_{0.5}$Ca$_{1.5}$MnO$_4$ and Pr(Sr$_{0.1}$Ca$_{0.9}$)$_2$Mn$_2$O$_7$, 
CO-OO produces clear anisotropy of electronic states, typically manifested by the 
optical conductivity spectra, in which the oscillator strength  at low energy 
region is more suppressed in orbital-stripe direction 
than in orbital zig-zag chain direction\cite{Tokunaga2008b,Tokunaga2006}

\section{Experiments and analyses}

We have performed structure analysis for single crystals of Eu$_{0.5}$Ca$_{1.5}$MnO$_4$ and 
Pr(Sr$_{0.1}$Ca$_{0.9}$)$_2$Mn$_2$O$_7$ grown by the floating zone method. 
The crystals were crashed into small grains. 
X-ray diffraction experiments were performed for twin-free single crystals with a diameter of about 30 $\mu$m 
on the beamline BL-1A at Photon Factory in KEK, Japan. 
The photon energy of the incident x-rays was tuned at 18 keV($\lambda$=0.688 \AA). 
X-ray beams were shaped into a square with the size of 300 $\mu$m $\times$ 300 $\mu$m by a collimator, 
which is enough larger than the size of samples. 
To detect x-rays, a large cylindrical imaging plate was utilized. 
Temperature was controlled by a nitrogen gas stream cryostat. 
The intensity data were operated to the F-tables by using the program of Rapid-Auto, Rigaku Corp. and MSC. 
The Sir2004 program\cite{Sir2004} was employed for the direct method. 
We used the program of CrystalStructure of Rigaku Corp. and MSC. for analyzing the crystal structure from 
the F-table. 
Absorption effects were not corrected, because the $\mu r$ were enough small in each sample. 
For Eu$_{0.5}$Ca$_{1.5}$MnO$_4$ and Pr(Sr$_{0.1}$Ca$_{0.9}$)$_2$Mn$_2$O$_7$, 
$\mu r$ are 0.40($\mu$=134.511 cm$^{-1}$) and 0.44($\mu$=146.027 cm$^{-1}$), respectively. 
To check this assumption, we tried to correct the absorption effects. 
No difference was observed for the structural data obtained with and without the absorption correction. 

\begin{figure}
\includegraphics*[width=85mm,clip]{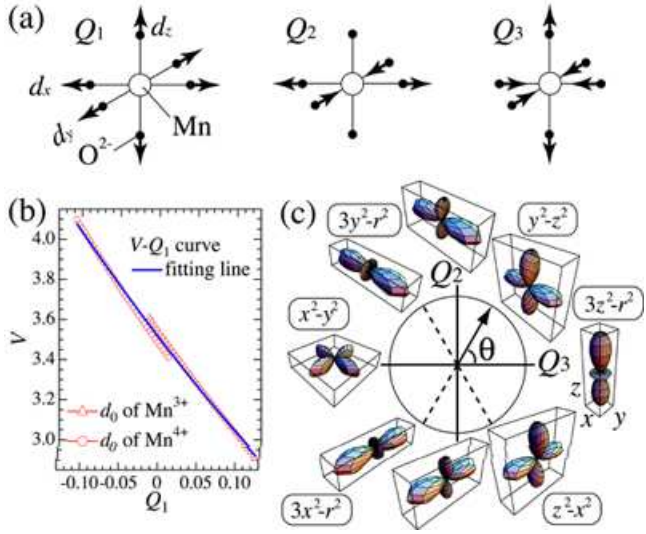}
\caption{\label{fig_1} (Color online) 
(a) Relevant distortion modes, $Q_1$, $Q_2$, and $Q_3$ of a MnO$_6$ octahedron. 
Arrows indicate shifts of O$^{2-}$ ions. 
(b) $V$-$Q_1$ curves calculated with using the bond valence sum formula. 
Triangles (circles) show the relation calculated for the $d_0$ value appropriate for an integer valence of 
Mn$^{3+}$ (Mn$^{4+}$). 
A solid line indicates a quadratic fitting with the values of triangles and circles. 
(c) Preferred orbital shapes of an $e_{g}$ electron with respect to the $Q_2$-$Q_3$ plane 
(Kanamori representation\cite{Kanamori1960}). 
}
\end{figure}

For the analysis, we used thus determined crystal structure data for Eu$_{0.5}$Ca$_{1.5}$MnO$_4$ and 
Pr(Sr$_{0.1}$Ca$_{0.9}$)$_2$Mn$_2$O$_7$, and the published data for Pr$_{0.5}$Ca$_{0.5}$MnO$_3$ 
by Goff \textit{et al.}\cite{Goff2004}. 
We adopt the localized-orbital picture, assuming \textit{a priori} 
the strong electron-lattice interaction\cite{footnote-2Dim}. 
Then, the breathing and Jahn-Teller distortion modes, $Q_1$, $Q_2$, and $Q_3$, 
can be related with charge disproportionation and orbital shape\cite{Kanamori1960}, 
and defined as 
\begin{equation}
\left(\begin{array}{@{\,}c@{\,}}
        Q_1 \\
        Q_2 \\
        Q_3 \\
       \end{array} \right)=
      1/\sqrt{6} \left(
       \begin{array}{@{\,}ccc@{\,}}
        \sqrt{2} & \sqrt{2} & \sqrt{2} \\
        \sqrt{3} & -\sqrt{3} & 0 \\
       -1 & -1 & 2 \\
       \end{array}
       \right)
\left(\begin{array}{@{\,}c@{\,}}
        d_{x}-\bar{d} \\
        d_{y}-\bar{d} \\
        d_{z}-\bar{d} \\
       \end{array} \right), 
\end{equation}
where $\mathit{d}_{x}$, $\mathit{d}_{y}$, and $\mathit{d}_{z}$ are bond lengths between Mn and O ions 
along the $x$, $y$, and $z$-axes, respectively, 
which are shown as schematic views of distorted MnO$_6$ octahedra in Fig. \ref{fig_1} (a). 
$\mathit{\bar{d}}$=1.956(2) \AA\ is the average bond length for Mn$^{3.5+}$(Ref.~\onlinecite{footnote2}). 
The approximate valences of Mn sites can be calculated from the bond valence sum\cite{Brown1985}, 
given by $V=\sum_{i} \exp{\left((d_0-d_{i})/B\right)}$. 
Here, $V$ is the calculated valence, $d_{i}$ is the $i$-th Mn-O bond length, 
$d_0$=1.760 \AA\ for Mn$^{3+}$, and 1.753 \AA\ for Mn$^{4+}$, and $B\sim$0.37 \AA\cite{Brown1985}. 
Basically, the bond valence sum is appropriate for ions with formal valence of integer such as Mn$^{3+}$ and Mn$^{4+}$ ions. 
To estimate the bond valence sum of intermediate valence states, we use a quadratic fit with 
the bond valence sum curves for Mn$^{3+}$ and Mn$^{4+}$, as shown in Fig. \ref{fig_1} (b)\cite{footnote3}. 
On the other hand, it has been well known as Kanamori representation \cite{Kanamori1960} 
that the orbital shape is related with the Jahn-Teller $Q_2$ and $Q_3$ modes, as shown in Fig. \ref{fig_1} (c). 
In the $Q_2$-$Q_3$ plane, the orbital state is thus described as 
$|d_{\theta}\rangle$=$\cos{(\frac{\theta}{2})} |d_{3z^2-r^2}\rangle + \sin{(\frac{\theta}{2})} |d_{x^2-y^2}\rangle$. 
Similar analysis based on the bond valence sum and the Kanamori representation has been applied to 
manganites in some literatures\cite{Goff2004,Carvajal1998,Zhou2006,Cussen2001}. 

\section{Results}

\begin{figure}
\includegraphics*[width=85mm,clip]{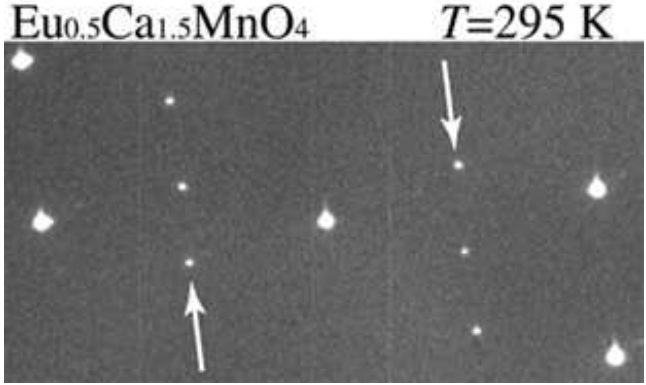}
\caption{\label{fig_2} (Color online) 
Typical diffraction image of Eu$_{0.5}$Ca$_{1.5}$MnO$_4$ in the CO-OO phase (295 K). 
The arrows indicate the superlattice reflections, whose intensities are 3 orders of magnitude weaker than 
those of the bright fundamental spots. 
}
\end{figure}
\begin{figure}
\includegraphics*[width=60mm,clip]{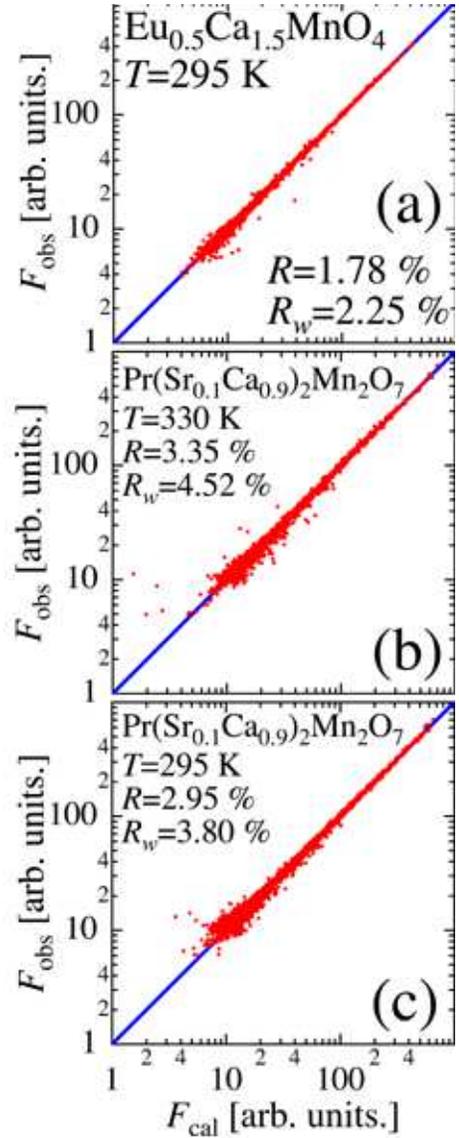}
\caption{\label{fig_3} (Color online) 
The comparison between observed ($F_{\mathrm{obs}}$) and calculated ($F_{\mathrm{cal}}$) structure factor of 
(a) Eu$_{0.5}$Ca$_{1.5}$MnO$_4$ at 295 K, 
(b) CO1 phase (330 K) of Pr(Sr$_{0.1}$Ca$_{0.9}$)$_2$Mn$_2$O$_7$, 
and (c) CO2 phase (295 K) of Pr(Sr$_{0.1}$Ca$_{0.9}$)$_2$Mn$_2$O$_7$, respectively. 
}
\end{figure}

In Fig. \ref{fig_2}, we show a typical diffraction pattern of Eu$_{0.5}$Ca$_{1.5}$MnO$_4$ in the CO-OO phase. 
Besides the fundamental Bragg spots, superlattice spots (indicated by arrows) due to cooperative Jahn-Teller distortion 
are clearly observed. 
The intensities of the superlattice reflections are three orders of magnitude weaker than those of the fundamental Bragg spots. 
There was no diffuse scattering intensity discerned around the superlattice spots, indicating the minimal effect of quenched 
disorder\cite{footnote1}. 
In Figs. \ref{fig_3} (a), (b), and (c), observed structure factor ($F_{\mathrm{obs}}$) is plotted against calculated one 
($F_{\mathrm{cal}}$) 
for Eu$_{0.5}$Ca$_{1.5}$MnO$_4$ at 295 K, for Pr(Sr$_{0.1}$Ca$_{0.9}$)$_2$Mn$_2$O$_7$ at 330 K, and at 295 K, respectively. 
The obtained reliability factors are $R$=1.78 \%, $R_w$=2.25 \% at CO-OO phase of Eu$_{0.5}$Ca$_{1.5}$MnO$_4$, 
$R$=3.35 \%, $R_w$=4.52 \% at CO1 phase ($T_{\mathrm{CO2}}\leqq T\leqq T_{\mathrm{CO1}}$), 
and $R$=2.95 \%, $R_w$=3.80 \% at CO2 phase ($T\leqq T_{\mathrm{CO2}}$) of 
Pr(Sr$_{0.1}$Ca$_{0.9}$)$_2$Mn$_2$O$_7$, respectively. 
Detailed crystal structural data are listed in APPENDIX, and CIF-files are available at elsewhere\cite{Okuyama2008}. 

\begin{figure*}
\includegraphics*[width=180mm,clip]{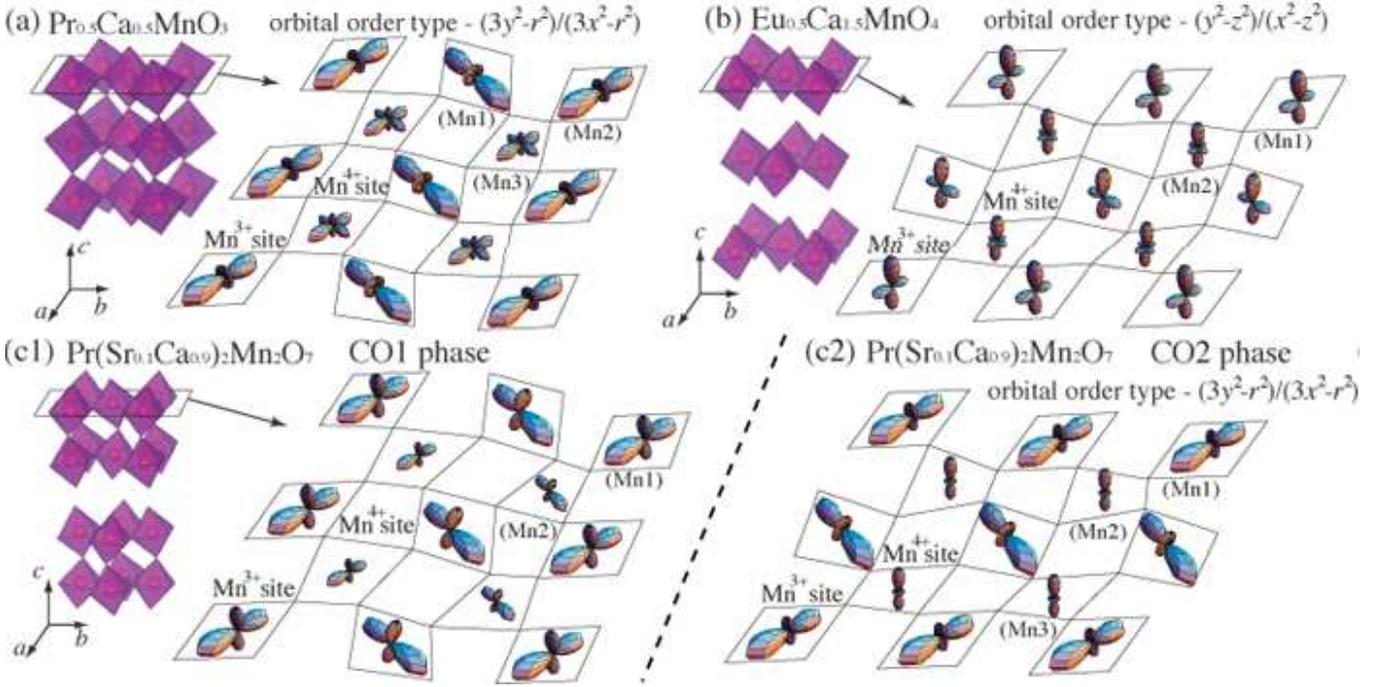}
\caption{\label{fig_4} (Color online) 
The CO-OO states schematically displayed for (a) Pr$_{0.5}$Ca$_{0.5}$MnO$_3$, 
(b) Eu$_{0.5}$Ca$_{1.5}$MnO$_4$, 
and (c1) CO1 phase (330 K) and (c2) CO2 phase (295 K) of Pr(Sr$_{0.1}$Ca$_{0.9}$)$_2$Mn$_2$O$_7$. 
The crystal structural data reported by Goff \textit{et al.}\cite{Goff2004} were used for the analysis of 
Pr$_{0.5}$Ca$_{0.5}$MnO$_3$. 
}
\end{figure*}

\begin{figure}
\includegraphics*[width=85mm,clip]{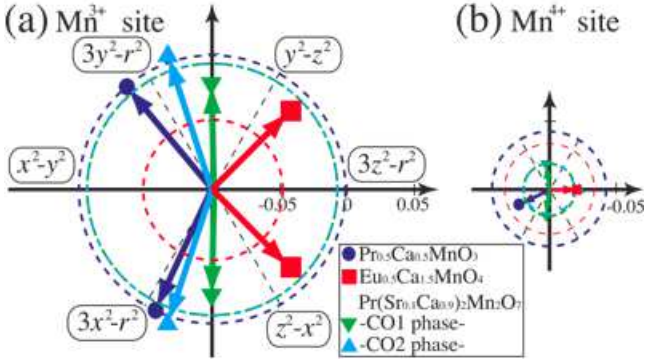}
\caption{\label{fig_5} (Color Online) 
Orbital shapes and charge states of (a) Mn$^{3+}$ and (b) Mn$^{4+}$ sites.
The radius of the circle is related to the Mn valence. 
The directions of arrows indicate the corresponding orbital shape. 
}
\end{figure}

Figure \ref{fig_4} (a) shows a schematic view of the CO-OO state of Pr$_{0.5}$Ca$_{0.5}$MnO$_3$, which is based on 
the crystal structural data by Goff \textit{et al.}\cite{Goff2004}, in which 
the space group and the lattice parameters at 10 K are reported to be $P2_1/m$ (No.11), $a$=5.43499(3) \AA, 
$b$=10.8700(2) \AA, and $c$=7.488923(6) \AA, respectively. 
Orbital stripes appear along the $a$-axis, which is parallel to the diagonal-glide plane of the orbital disordered phase. 
The orbital shapes at Mn$^{3+}$ and Mn$^{4+}$ sites of Pr$_{0.5}$Ca$_{0.5}$MnO$_3$ obtained by the present analysis 
are indicated in Figs. \ref{fig_5} (a) and (b), respectively, 
in which the radius of the circle is related to $Q_1$ [$Q_1$=-0.1 (0.1) corresponds to $V\sim$+4 (+3)], and  
the direction of each arrow indicates the orbital state $|d_{\theta}\rangle$ calculated from $Q_2$ and $Q_3$. 
There are two crystallographically inequivalent Mn$^{3+}$ sites termed Mn1 and Mn2, 
and one Mn$^{4+}$ site (Mn3) in the CO-OO phase. 
The obtained values of $d_x$, $d_y$, $d_z$, $Q_2$, $Q_3$, and $V$ are 
listed in Table \ref{table_1}\cite{footnote-Vvalue}. 
From these values, charge disproportionation between nominal Mn$^{3+}$ and Mn$^{4+}$ sites 
is estimated to be roughly 22\%, 
which is very close to the value reported for Nd$_{0.5}$Sr$_{0.5}$MnO$_3$ using resonant X-ray scattering 
technique\cite{Martin2004}. 
Furthermore, for the Mn1 and Mn2 sites, the ($3y^2-r^2$)/($3x^2-r^2$)-type orbital shapes are obtained. 
For the Mn3 site, by contrast, the $Q_2$ and $Q_3$ values are small as compared to $Q_1$ value, 
indicating the least Jahn-Teller distortion or unlifted orbital degeneracy with almost isotropic electron density, 
namely, $\rho(\boldsymbol{r})\propto\lambda|\Psi_{3z^2-r^2}|^2+|\Psi_{x^2-y^2}|^2$, with $\lambda$ slightly smaller than 1. 

\begin{table}
\caption{\label{table_1}
Distortion of MnO$_6$ octahedra in Pr$_{0.5}$Ca$_{0.5}$MnO$_3$ at 10 K\cite{Goff2004}. 
$d_i$ and $Q_j$ ($i$=$x$, $y$, $z$; $j$=2, 3) are in unit of \AA. 
The space group is $P2_1/m$ (No.11), 
and the lattice parameters are $a$=5.43499(3) \AA, $b$=10.8700(2) \AA, and $c$=7.488923(6) \AA, respectively.
}
\begin{ruledtabular}
\begin{tabular}{ccccccc}
 &$d_x$&$d_y$&$d_z$&$Q_2$&$Q_3$&$V$\\
\hline
Mn1 & 2.039(6) & 1.930(5) & 1.9061(6) & 0.077(4) & -0.064(5) & 3.48(2)\\
Mn2 & 1.911(8) & 2.027(5) & 1.9216(3) & -0.082(5) & -0.039(6) & 3.53(3)\\
Mn3 & 1.933(5) & 1.948(6) & 1.914(3) & -0.011(6) & -0.021(6) & 3.72(4)\\
\end{tabular}
\end{ruledtabular}
\end{table}

The CO-OO state of Eu$_{0.5}$Ca$_{1.5}$MnO$_4$ with single-layer Mn-O network is schematically illustrated in 
Fig. \ref{fig_4} (b). 
The crystal structural analysis showed that the space group is $Pmnb$ (No.62) of orthorhombic crystal system, and 
the lattice parameters are $a$=10.6819(7) \AA, $b$=5.4071(3) \AA, and $c$=11.7018(11) \AA \ at 295 K, respectively. 
Orbital stripes line up along the $b$-axis, which is perpendicular to the diagonal-glide plane, 
giving rise to the observed optical anisotropy~\cite{Tokunaga2008b}.
The obtained parameters are listed in Table \ref{table_2}. 
About 8\% charge disproportionation between Mn$^{3+}$ and Mn$^{4+}$ sites is observed, 
which should be interpreted as indicating that the actual charge disproportionation is almost negligible, 
taking into account the semi-quantitative nature of the bond valence sum analysis. 
The negligible charge disproportionation is in excellent accord with the theoretical prediction\cite{Mahadevan2001}
for the single-layer compound  La$_{0.5}$Sr$_{1.5}$MnO$_4$. 
For Mn1 site, the ($y^2-z^2$)/($x^2-z^2$)-type orbital shape is observed, as shown in Fig. \ref{fig_5}. 
This orbital shape is also consistent with the result  of x-ray linear dichroism experiment\cite{Huang2004}. 
In contrast, for Mn2 site, $Q_2$ and $Q_3$ values are small with respect to $Q_1$ value, 
indicating almost isotropic electron density $\rho(\boldsymbol{r})\propto\lambda|\Psi_{3z^2-r^2}|^2+|\Psi_{x^2-y^2}|^2$, 
with $\lambda$ slightly larger than 1. 
Therefore, Mn1 and Mn2 sites can be regarded as nominal Mn$^{3+}$ and Mn$^{4+}$ site in the light of orbital activity, 
albeit minimal charge disproportionation. 

\begin{table}
\caption{\label{table_2}
Distortion of MnO$_6$ octahedra in Eu$_{0.5}$Ca$_{1.5}$MnO$_4$ at 295 K. 
The space group is $Pmnb$ (No.62) of orthorhombic crystal system, and the lattice parameters are 
$a$=10.6819(7) \AA, $b$=5.4071(3) \AA, and $c$=11.7018(11) \AA, respectively.
}
\begin{ruledtabular}
\begin{tabular}{ccccccc}
 &$d_x$&$d_y$&$d_z$&$Q_2$&$Q_3$&$V$\\
\hline
Mn1 & 1.943(4) & 1.863(3) & 1.972(5) & 0.057(3) & 0.056(3) & 3.77(2)\\
Mn2 & 1.910(5) & 1.910(5) & 1.934(4) & 0.000(3) & 0.019(3) & 3.85(2)\\
\end{tabular}
\end{ruledtabular}
\end{table}

The pseudo cubic and single-layer compounds show clear contrast in orbital shape and charge 
disproportionation; 22\% charge disproportionation and ($3y^2-r^2$)/($3x^2-r^2$)-type orbital shape 
of Mn$^{3+}$ site for pseudo cubic, while negligibly small charge disproportionation and 
($y^2-z^2$)/($x^2-z^2$)-type orbital shape for single-layer. 
From the crystallographic point of view, the double-layer system has an intermediate structure between pseudo cubic 
and single-layer. 
We show in Fig. \ref{fig_4} (c) schematic pictures of the CO-OO states of 
Pr(Sr$_{0.1}$Ca$_{0.9}$)$_2$Mn$_2$O$_7$ with double-layer Mn-O network 
based on the structural parameters listed in Table \ref{table_3}. 
The space groups are $Pbnm$ (No. 62) with $a$=5.4087(2) \AA, $b$=10.9171(5) \AA, and $c$=19.2312(12) \AA \ 
at 330 K (CO1) and $Am2m$ (No. 38) with $a$=10.8026(7) \AA, $b$=5.4719(4) \AA, and $c$=19.2090(10) \AA \ 
at 295 K (CO2), respectively. 
In the CO1 phase, the orbital stripes along the $a$-axis are clearly seen, which is parallel to the diagonal-glide plane. 
We observe about 39\% charge disproportionation in the CO1 phase. 
For Mn1 site, the orbital shape is of intermediate type between ($3y^2-r^2$)/($3x^2-r^2$) and 
($y^2-z^2$)/($z^2-x^2$), and the similar orbital shape is obtained for Mn2 site (Fig. \ref{fig_5}). 
Pr(Sr$_{0.1}$Ca$_{0.9}$)$_2$Mn$_2$O$_7$ undergoes a transition with 90-degree rotation of orbital stripes 
at $T_{\mathrm{CO2}}$$\sim$315 K\cite{Tokunaga2006}. 
In the CO2 phase, the orbital stripes run along the $b$-axis, as clearly seen in Fig. \ref{fig_4} (c2), in accord with 
the previous report\cite{Tokunaga2006}. 
There are one Mn$^{3+}$ site and two crystallographically inequivalent Mn$^{4+}$ sites, termed respectively 
Mn1, Mn2, and Mn3, in the CO2 phase. 
The obtained charge disproportionation of about 39\% is comparable with that of CO1 phase, 
whereas the ($3y^2-r^2$)/($3x^2-r^2$)-type orbital shape as in Pr$_{0.5}$Ca$_{0.5}$MnO$_3$ is suggested for the 
Mn$^{3+}$ sites, as shown in Fig. \ref{fig_5}.   In both phases, the observed direction of orbital stripes is 
consistent with the optical anisotropy~\cite{Tokunaga2006}.

\begin{table}
\caption{\label{table_3}
Distortion of MnO$_6$ octahedra at CO1 phase (330 K) and CO2 phase (295 K) of 
Pr(Sr$_{0.1}$Ca$_{0.9}$)$_2$Mn$_2$O$_7$. 
The space group is $Pbnm$ (No. 62) and $Am2m$ (No. 38) of orthorhombic crystal system 
at 330 K and 295 K, respectively, and the lattice parameters are 
$a$=5.4087(2) \AA, $b$=10.9171(5) \AA, and $c$=19.2312(12) \AA \ at 330 K, and
$a$=10.8026(7) \AA, $b$=5.4719(4) \AA, and $c$=19.2090(10) \AA \ at 295 K, respectively.
}
\begin{ruledtabular}
\begin{tabular}{ccccccc}
330 K &$d_x$&$d_y$&$d_z$&$Q_2$&$Q_3$&$V$\\
\hline
Mn1 & 2.007(4) & 1.899(3) & 1.952(4) & 0.076(3) & -0.001(3) & 3.53(2)\\
Mn2 & 1.898(4) & 1.925(5) & 1.908(4) & 0.019(3) & -0.003(3) & 3.92(2)\\
\end{tabular}
\begin{tabular}{ccccccccc}
295 K &$d_x$&$d_y$&$d_z$&$Q_2$&$Q_3$&$V$\\
\hline
Mn1 & 2.032(2) & 1.898(2) & 1.926(3) & 0.095(2) & -0.032(2) & 3.53(2)\\
Mn2 & 1.901(2) & 1.901(2) & 1.929(3) & 0.000(2) & 0.023(2) & 3.92(2)\\
Mn3 & 1.905(2) & 1.905(2) & 1.924(3) & 0.000(2) & 0.016(2) & 3.91(2)\\
\end{tabular}
\end{ruledtabular}
\end{table}

\section{Discussion}

While the stripe-type orbital pattern commonly observed in the three materials gives rise to the same
CE-type magnetic ordering,  the orbital shape does depend on the structural difference, and
seems to be governed by the difference in the stacking-sequence of (rare-earth/alkaline-earth)-oxygen (AO) 
planes and MnO$_2$ planes. 
Note that the AO plane contains apical oxygens of MnO$_6$ octahedra. 
In the charge- and orbital-disordered phase of Pr$_{0.5}$Ca$_{0.5}$MnO$_3$, 
MnO$_6$ octahedron is almost isotropic with slight compression along $c$-axis
due to the pseudo cubic symmetry of the crystal structure\cite{Jirak2000}. 
Below the CO-OO transition, $e_g$ electrons tend to maximize the kinetic energy gain due to the
local double exchange interaction with the neighboring Mn$^{4+}$ $t_{2g}$ spins on the $ab$-plane,
and the ($3y^2-r^2$)/($3x^2-r^2$)-type orbital shape is favored.
In Eu$_{0.5}$Ca$_{1.5}$MnO$_4$, by contrast, the bond length between Mn and apical oxygen 
is longer than that within the plane\cite{Okuyama2008} even in the charge- and orbital-disordered phase at 360 K.
This is because a single MnO$_2$ plane is negatively charged, and hence apical oxygen and rare-earth/alkaline-earth ions 
on the adjacent layers tend to be apart from and close to the MnO$_2$ plane, respectively. 
The body-centered nature of the K$_2$NiF$_4$-like structure allows this type of local lattice distortion 
in a cooperative manner. (It should be noted that this lattice distortion is not driven by the Jahn-Teller interaction, but 
due to the purely lattice structural effect, as evidenced by the presence of similar lattice distortion in  
La$_2$NiO$_4$\cite{Jorgensen1989} in which Ni$^{2+}$ (with two $e_g$ electrons) with $S$=1 is Jahn-Teller inactive.)
Therefore, the electron orbital extended along the $c$-axis is stabilized by such a distortion at high temperatures. 
As the temperature is lowered, however, $e_g$ electrons tend to favor the orbital shape extended within the MnO$_2$ planes 
so as to gain the spin exchange energy, similarly to the case of  Pr$_{0.5}$Ca$_{0.5}$MnO$_3$,  
thereby establishing ($y^2-z^2$)/($z^2-x^2$)-type orbital order at $T_{\mathrm{CO}}$. 
The structure of Pr(Sr$_{0.1}$Ca$_{0.9}$)$_2$Mn$_2$O$_7$ compound can be viewed as intermediate
between the single layer and infinite layer compounds.
Therefore, it is likely that an intermediate orbital state between Pr$_{0.5}$Ca$_{0.5}$MnO$_3$ 
and Eu$_{0.5}$Ca$_{1.5}$MnO$_4$ 
is favored in the intermediate CO1 phase. 
In the CO2 phase, however, the ($3y^2-r^2$)/($3x^2-r^2$)-type orbital shape is realized as in the pseudo cubic case, 
in which the magnetic interaction as favoring the CE-type order may play a role. 

As the layer number is increased, charge disproportionation increases from the smallest value of 8\%  
for the single layer compound to  39 \% for the double layer material, and then rather decreases to 22 \%
for the infinite layer compound.
In the single layer system, the development of order parameter may be suppressed due to the enhanced fluctuation effect.
In  the double layer system, fluctuation effect would be reduced, and the larger charge disproportionation is observed.
The reason why the observed charge disproportionation is smaller in the infinite layer material than in
double layer compound is not clear at present, but the enhanced double exchange interaction
in the infinite layer material due to the increased dimensionality
might be the origin.

\section{Summary}

In summary, we have revealed that the charge disproportionation between nominal Mn$^{3+}$ and Mn$^{4+}$ is in reality 
much smaller than unity, and in particular, negligible in single-layer system. 
This result is in contrast with the belief at the early stage of the research of the CO-OO phenomena, 
but in accord with the  recent results
\cite{Brink1999,Mahadevan2001,Subias1997,Garcia2001,Popovic2002,Martin2004,Goff2004,Loudon2005,Milward2005,Cox2008}.
This indicates that various interactions, such as electron-lattice,  Coulomb repulsion, and  magnetic interactions, 
should  be taken into account to correctly understand the nature of charge-ordered state. 
The orbital shape is different among the three compounds despite the same symmetry of the CO-OO pattern 
in a single MnO$_2$ plane, structurally confirming the previous conclusions based on different experimental 
techniques\cite{Radaelli1997,Huang2004}. 
The charge disproportionation and orbital shape in these materials are dominated by the dimension of Mn-O network, and the 
local environment of the apical oxygen seems to be particularly important. 

\begin{acknowledgments}
The authors thank S. Ishiwata and Y. Tomioka for fruitful discussions. 
This study was performed with the approval of the Photon Factory Program Advisory Committee (No.2006S2-005). 
\end{acknowledgments}

\appendix

\section{Crystal-structural data}

We present the crystal structural data of Eu$_{0.5}$Ca$_{1.5}$MnO$_4$ and 
Pr(Sr$_{0.1}$Ca$_{0.9}$)$_2$Mn$_2$O$_7$. 
Table \ref{table_4} shows the distortion parameters $d_x$, $d_y$, $d_z$, $Q_1$, $Q_2$, $Q_3$, $V$, 
and $\theta$ of Pr$_{0.5}$Ca$_{0.5}$MnO$_3$ calculated from the crystallographic data of 
Goff \textit{et al.}\cite{Goff2004}. 
The crystal structural data at 360 K (disordered phase) and 295 K (CO-OO phase) of 
Eu$_{0.5}$Ca$_{1.5}$MnO$_4$ are shown in Table \ref{table_5} and Fig. \ref{table_6}, respectively. 
From these crystal data, the distortion parameters of 360 K and 295 K are calculated and 
summarized as Table \ref{table_7}. 
For Pr(Sr$_{0.1}$Ca$_{0.9}$)$_2$Mn$_2$O$_7$, 
the crystal structure data in disordered phase (at 405 K) are presented in Table \ref{table_8}. 
For two CO-OO phases, the data at 330 K in CO1 phase 
($T_{\mathrm{CO2}}\leqq T\leqq T_{\mathrm{CO1}}$) and at 295 K in CO2 ($T\leqq T_{\mathrm{CO2}}$) 
are listed in Fig. \ref{table_9} and Fig. \ref{table_10}, respectively. 
In Table \ref{table_11}, the distortion parameters at 405 K, 330 K, and 295 K are reported. 
In these crystal structural data, $x$, $y$, and $z$ indicate the fractional coordinates. 
$g$ is the site-occupation number. 
Anisotropic atomic displacement parameters are represented as $U_{11}$, $U_{22}$, $U_{33}$, $U_{12}$, $U_{13}$, 
and $U_{23}$, while $B_j$ and $\langle u_j^2\rangle$ ($B_j$=8$\pi^2$$U_j$=8$\pi^2$$\langle u_j^2\rangle$) 
are the isotropic atomic displacement parameter and the mean square atomic displacement of the ion $j$, respectively. 
From the $B_j$, the Debye-Waller factor is expressed as $\exp{(-B_j(\sin{\theta_K}/\lambda)^2)}$, 
where $\theta_K$ and $\lambda$ are the Bragg angle and wave length, respectively.

\begin{table*}
\caption{\label{table_4}
Distortion of MnO$_6$ octahedra in Pr$_{0.5}$Ca$_{0.5}$MnO$_3$ at 10 K\cite{Goff2004}. 
}
\begin{ruledtabular}
\begin{tabular}{ccccccccc}
 &$d_x$ (\AA$^2$)&$d_y$ (\AA$^2$)&$d_z$ (\AA$^2$)&$Q_1$&$Q_2$&$Q_3$&$V$&$\theta$\\
\hline
Mn1 & 2.039(6) & 1.930(5) & 1.9061(6) & 0.004(4) & 0.077(4) & -0.064(5) & 3.48(2) & 130(4)$^{\circ}$\\
Mn2 & 1.911(8) & 2.027(5) & 1.9216(3) & -0.005(5) & -0.082(5) & -0.039(6) & 3.53(3) & 245(5)$^{\circ}$\\
Mn3 & 1.933(5) & 1.948(6) & 1.914(3) & -0.042(6) & -0.011(6) & -0.021(6) & 3.72(4) & 21(2)$\times$10$^{\circ}$\\
\end{tabular}
\end{ruledtabular}
\end{table*}
\begin{table*}
\caption{\label{table_5}
The structure parameters of Eu$_{0.5}$Ca$_{1.5}$MnO$_4$ in the disordered phase at 360 K. 
The 3048 reflections were observed, and 966 of them are independent. 
The 29 variables were used for the refinement. 
The space group is $Bmab$ (No. 64) of orthorhombic crystal system. 
The latticce parameters are as follows: $a$=5.3638(19) \AA, $b$=5.4088(9) \AA, $c$=11.738(2) \AA. 
The reliability factors are $R$= 2.40\%, $R_{w}$=2.55\%, GOF(Goodness of fit)=1.062. 
}
\begin{ruledtabular}
\begin{tabular}{ccccccc}
 &site&$x$&$y$&$z$&$g$&$B$ (\AA$^2$)\\
\hline
Eu1 & 8$f$ & 0 & 0.9933(4) & 0.64219(16) & 1/2 & 0.645(15)\\
Ca1 & 8$f$ & 0 & 0.9854(5) & 0.6427(2) & 3/2 & 0.75(2)\\
Mn & 4$a$ & 0 & 0 & 0 & 1 & 0.412(5)\\
O1 & 8$e$ & 1/4 & 3/4 & 0.98777(8) & 1 & 1.097(19)\\
O2 & 8$f$ & 0 & 0.9623(2) & 0.16570(7) & 1 & 1.51(2)\\
\end{tabular}
\begin{tabular}{ccccccc}
 &$U_{11}$ (\AA$^2$)&$U_{22}$ (\AA$^2$)&$U_{33}$ (\AA$^2$)&$U_{12}$ (\AA$^2$)&$U_{13}$ (\AA$^2$)&
 $U_{23}$ (\AA$^2$)\\
\hline
Eu1 & 0.0111(8) & 0.00898(17) & 0.0044(2) & 0.0000 & 0.0000 & 0.00132(14)\\
Ca1 & 0.0129(9) & 0.0097(2) & 0.0060(3) & 0.0000 & 0.0000 & -0.00106(19)\\
Mn & 0.0056(2) & 0.00414(4) & 0.00593(6) & 0.0000 & 0.0000 & -0.00030(2)\\
O1 & 0.0111(10) & 0.01142(19) & 0.0192(2) & 0.0028(3) & 0.0000 & 0.0000\\
O2 & 0.0314(11) & 0.0202(3) & 0.00583(15) & 0.0000 & 0.0000 & -0.00107(19)\\
\end{tabular}
\end{ruledtabular}
\end{table*}

\begin{table*}
\caption{\label{table_6}
The structure parameters of Eu$_{0.5}$Ca$_{1.5}$MnO$_4$ in the charge- and orbital-ordered phase at 295 K. 
The 13758 reflections were observed, and 5552 of them are independent. 
The 95 variables were used for the refinement. 
The space group is $Pmnb$ (No. 62) of orthorhombic crystal system. 
The latticce parameters are as follows: $a$=10.6819(7) \AA, $b$=5.4071(3) \AA, $c$=11.7018(11) \AA. 
The reliability factors are $R$= 1.78\, $R_{w}$=2.25\, GOF=1.058. 
}
\begin{ruledtabular}
\begin{tabular}{ccccccc}
 &site&$x$&$y$&$z$&$g$&$B$ (\AA$^2$)\\
\hline
Eu1 & 4$c$ & 3/4 & 0.004300(10) & 0.143400(10) & 0.25 & 0.513(16)\\
Eu2 & 8$d$ & 0.49964(2) & 0.00800(3) & 0.35889(3) & 0.25 & 0.56(2)\\
Eu3 & 4$c$ & 3/4 & 0.504900(10) & 0.642900(10) & 0.25 & 0.370(19)\\
Ca1 & 4$c$ & 3/4 & -0.016400(10) & 0.140700(10) & 0.75 & 0.28(2)\\
Ca2 & 8$d$ & 0.49916(3) & 0.01430(3) & 0.35555(3) & 0.75 & 0.84(3)\\
Ca3 & 4$c$ & 3/4 & 0.530700(10) & 0.641700(10) & 0.75 & 0.417(11)\\
Mn1 & 4$a$ & 0 & 1/2 & 1/2 & 1 & 0.354(14)\\
Mn2 & 4$c$ & 3/4 & 0.00604(5) & 0.49926(10) & 1 & 0.344(15)\\
O1 & 8$d$ & 0.8753(3) & 0.2610(5) & 0.5134(2) & 1 & 0.91(2)\\
O2 & 8$d$ & 0.6259(4) & 0.7579(7) & 0.4874(2) & 1 & 1.02(2)\\
O3 & 4$c$ & 3/4 & 0.0503(10) & 0.3359(3) & 1 & 1.02(4)\\
O4 & 4$c$ & 3/4 & 0.9599(13) & 0.6637(3) & 1 & 1.07(4)\\
O5 & 8$d$ & 0.9992(2) & 0.5337(13) & 0.6678(4) & 1 & 1.62(5)\\
\end{tabular}
\begin{tabular}{ccccccc}
 &$U_{11}$ (\AA$^2$)&$U_{22}$ (\AA$^2$)&$U_{33}$ (\AA$^2$)&$U_{12}$ (\AA$^2$)&$U_{13}$ (\AA$^2$)&
 $U_{23}$ (\AA$^2$)\\
\hline
Eu1 & 0.0088(4) & 0.00441(17) & 0.0063(5) & 0.0000 & 0.0000 & -0.0010(2)\\
Eu2 & 0.0060(5) & 0.0118(6) & 0.0034(6) & -0.00010(18) & -0.0010(3) & 0.0018(2)\\
Eu3 & 0.0076(6) & 0.0011(2) & 0.0053(6) & 0.0000 & 0.0000 & 0.0004(2)\\
Ca1 & 0.0056(5) & 0.0015(2) & 0.0033(7) & 0.0000 & 0.0000 & -0.0014(2)\\
Ca2 & 0.0158(5) & 0.0102(7) & 0.0059(8) & 0.0011(2) & 0.0010(3) & -0.0003(4)\\
Ca3 & 0.0087(3) & 0.00131(19) & 0.0059(4) & 0.0000 & 0.0000 & 0.00096(17)\\
Mn1 & 0.0017(2) & 0.00518(19) & 0.0065(6) & -0.00023(5) & -0.0001(2) & -0.0001(2)\\
Mn2 & 0.0053(3) & 0.00314(16) & 0.0046(6) & 0.0000 & 0.0000 & -0.0005(2)\\
O1 & 0.0080(3) & 0.0080(4) & 0.0184(10) & -0.0038(2) & 0.0027(6) & -0.0001(4)\\
O2 & 0.0092(3) & 0.0142(4) & 0.0154(10) & -0.0005(2) & 0.0031(8) & -0.0002(5)\\
O3 & 0.0239(11) & 0.0139(8) & 0.0010(12) & 0.0000 & 0.0000 & -0.0002(5)\\
O4 & 0.0222(12) & 0.0145(7) & 0.0040(16) & 0.0000 & 0.0000 & 0.0014(7)\\
O5 & 0.0325(14) & 0.0204(13) & 0.0087(15) & -0.0014(4) & -0.0012(5) & 0.0044(8)\\
\end{tabular}
\end{ruledtabular}
\end{table*}

\begin{table*}
\caption{\label{table_7}
Distortion of MnO$_6$ octahedra in Eu$_{0.5}$Ca$_{1.5}$MnO$_4$ at 360 K(disordered phase) and 
295 K(ordered phase). 
}
\begin{ruledtabular}
\begin{tabular}{ccccccccc}
360 K &$d_x$ (\AA$^2$)&$d_y$ (\AA$^2$)&$d_z$ (\AA$^2$)&$Q_1$&$Q_2$&$Q_3$&$V$&$\theta$\\
\hline
Mn1 & 1.9098(4) & 1.9098(4) & 1.9556(9) & -0.054(2) & 0.000(2) & 0.037(2) & 3.78(2) & 0(3)$^{\circ}$\\
\end{tabular}
\begin{tabular}{ccccccccc}
295 K &$d_x$ (\AA$^2$)&$d_y$ (\AA$^2$)&$d_z$ (\AA$^2$)&$Q_1$&$Q_2$&$Q_3$&$V$&$\theta$\\
\hline
Mn1 & 1.943(4) & 1.863(3) & 1.972(5) & -0.052(3) & 0.057(3) & 0.056(3) & 3.77(2) & 45(3)$^{\circ}$\\
Mn2 & 1.910(5) & 1.910(5) & 1.934(4) & -0.066(3) & 0.000(3) & 0.019(3) & 3.85(2) & 0(1)$\times$10$^{\circ}$\\
\end{tabular}
\end{ruledtabular}
\end{table*}
\begin{table*}
\caption{\label{table_8}
The structure parameters of Pr(Sr$_{0.1}$Ca$_{0.9}$)$_2$Mn$_2$O$_7$ in the disordered phase at 405 K. 
The 6074 reflections were observed, and 2088 of them are independent. 
The 48 variables were used for the refinement. 
The space group is $Amam$ (No. 63) of orthorhombic crystal system. 
The latticce parameters are as follows: $a$=5.4080(5) \AA, $b$=5.4599(5) \AA, $c$=19.266(3) \AA. 
The reliability factors are $R$= 3.42\%, $R_{w}$=4.31\%, GOF=1.088. 
}
\begin{ruledtabular}
\begin{tabular}{ccccccc}
 &site&$x$&$y$&$z$&$g$&$B$ (\AA$^2$)\\
\hline
Pr1 & 8$g$ & 3/4 & 0.2375(6) & 0.1844(2) & 0.21 & 0.86(2)\\
Pr2 & 4$c$ & 3/4 & 0.25011(18) & 0 & 0.58 & 0.763(16)\\
Sr1 & 8$g$ & 3/4 & 0.2413(5) & 0.1833(2) & 0.08 & 0.73(2)\\
Sr2 & 4$c$ & 3/4 & 0.2523(8) & 0 & 0.04 & 1.11(7)\\
Ca1 & 8$g$ & 3/4 & 0.2413(5) & 0.1833(2) & 0.71 & 0.73(2)\\
Ca2 & 4$c$ & 3/4 & 0.2523(8) & 0 & 0.38 & 1.11(7)\\
Mn & 8$g$ & 3/4 & 0.74812(2) & 0.09956(2) & 1 & 0.484(3)\\
O1 & 8$e$ & 0 & 1/2 & 0.10744(12) & 1 & 1.42(2)\\
O2 & 8$e$ & 0 & 0 & 0.08776(11) & 1 & 1.33(2)\\
O3 & 8$g$ & 3/4 & 0.7950(2) & 0.19846(11) & 1 & 1.37(2)\\
O4 & 4$c$ & 3/4 & 0.6950(4) & 0 & 1 & 1.52(4)\\
\end{tabular}
\begin{tabular}{ccccccc}
 &$U_{11}$ (\AA$^2$)&$U_{22}$ (\AA$^2$)&$U_{33}$ (\AA$^2$)&$U_{12}$ (\AA$^2$)&$U_{13}$ (\AA$^2$)&
 $U_{23}$ (\AA$^2$)\\
\hline
Pr1 & 0.0156(7) & 0.0111(4) & 0.0059(13) & 0.0000 & 0.0000 & 0.0006(5)\\
Pr2& 0.0126(3) & 0.0106(2) & 0.0059(6) & 0.0000 & 0.0000 & 0.0000\\
Sr1& 0.0087(10) & 0.0103(3) & 0.0087(3) & 0.0000 & 0.0000 & 0.0005(4)\\
Sr2& 0.0062(9) & 0.0096(10) & 0.026(3) & 0.0000 & 0.0000 & 0.0000\\
Ca1& 0.0087(10) & 0.0103(3) & 0.0087(3) & 0.0000 & 0.0000 & 0.0005(4)\\
Ca2& 0.0062(9) & 0.0096(10) & 0.026(3) & 0.0000 & 0.0000 & 0.0000\\
Mn& 0.00527(6) & 0.00597(5) & 0.00715(15) & 0.0000 & 0.0000 & -0.00005(4)\\
O1& 0.0150(4) & 0.0164(3) & 0.0226(11) & 0.0062(2) & 0.0000 & 0.0000\\
O2& 0.0158(4) & 0.0165(3) & 0.0182(11) & -0.0046(3) & 0.0000 & 0.0000\\
O3& 0.0263(6) & 0.0176(4) & 0.0080(10) & 0.0000 & 0.0000 & -0.0008(4)\\
O4& 0.0325(10) & 0.0164(5) & 0.0090(14) & 0.0000 & 0.0000 & 0.0000\\
\end{tabular}
\end{ruledtabular}
\end{table*}

\begin{table*}
\caption{\label{table_9}
The structure parameters of Pr(Sr$_{0.1}$Ca$_{0.9}$)$_2$Mn$_2$O$_7$ in the charge- and orbital-ordered phase 
(CO1)  at 330 K. 
The 25327 reflections were observed, and 8255 of them are independent. 
The 130 variables were used for the refinement. 
The space group is $Pbnm$ (No. 62) of orthorhombic crystal system. 
The latticce parameters are as follows: $a$=5.4087(2) \AA, $b$=10.9171(5) \AA, $c$=19.2312(12) \AA. 
The reliability factors are $R$= 3.35\, $R_{w}$=4.52\, GOF=0.976. 
}
\begin{ruledtabular}
\begin{tabular}{ccccccc}
 &site&$x$&$y$&$z$&$g$&$B$ (\AA$^2$)\\
\hline
Pr1 & 8$d$ & 0.75725(3) & 0.74371(3) & 0.56534(3) & 0.21 & 0.66(3)\\
Pr2 & 4$c$ & 0.260500(10) & 0.747320(10) & 1/4 & 0.58 & 0.643(9)\\
Pr3 & 4$c$ & 0.746400(10) & 0.498900(10) & 1/4 & 0.58 & 0.787(16)\\
Pr4 & 8$d$ & 0.74868(3) & 0.48804(3) & 0.06413(3) & 0.21 & 0.514(15)\\
Sr1 & 8$d$ & 0.76207(3) & 0.74477(3) & 0.56679(3) & 0.08 & 0.73(2)\\
Sr2 & 4$c$ & 0.255200(10) & 0.759300(10) & 1/4 & 0.04 & 0.48(3)\\
Sr3 & 4$c$ & 0.732000(10) & 0.502000(10) & 1/4 & 0.04 & 0.43(3)\\
Sr4 & 8$d$ & 0.74744(3) & 0.500610(3) & 0.06809(3) & 0.08 & 0.502(12)\\
Ca1 & 8$d$ & 0.76207(3) & 0.74477(3) & 0.56679(3) & 0.71 & 0.73(2)\\
Ca2 & 4$c$ & 0.255200(10) & 0.759300(10) & 1/4 & 0.38 & 0.48(3)\\
Ca3 & 4$c$ & 0.732000(10) & 0.502000(10) & 1/4 & 0.38 & 0.43(3)\\
Ca4 & 8$d$ & 0.74744(3) & 0.500610(3) & 0.06809(3) & 0.71 & 0.502(12)\\
Mn1 & 8$d$ & 0.24701(4) & 0.50147(13) & 0.34933(6) & 1 & 0.431(8)\\
Mn2 & 8$d$ & 0.75953(5) & 0.74929(12) & 0.34970(6) & 1 & 0.441(6)\\
O1 & 8$d$ & 0.0096(6) & 0.6250(3) & 0.36026(11) & 1 & 1.20(2)\\
O2 & 8$d$ & 0.5089(7) & 0.6264(3) & 0.35514(12) & 1 & 1.33(3)\\
O3 & 4$c$ & 0.7685(5) & 0.7174(4) & 1/4 & 1 & 1.06(4)\\
O4 & 4$c$ & 0.2397(4) & 0.5220(5) & 1/4 & 1 & 1.20(6)\\
O5 & 8$d$ & 0.4858(5) & 0.3725(3) & 0.33750(12) & 1 & 1.08(2)\\
O6 & 8$d$ & 0.9798(4) & 0.3707(2) & 0.33743(12) & 1 & 0.98(2)\\
O7 & 8$d$ & 0.7516(3) & 0.7682(5) & 0.4461(2) & 1 & 1.23(5)\\
O8 & 8$d$ & 0.2518(2) & 0.4731(4) & 0.4510(2) & 1 & 0.92(4)\\
\end{tabular}
\begin{tabular}{ccccccc}
 &$U_{11}$ (\AA$^2$)&$U_{22}$ (\AA$^2$)&$U_{33}$ (\AA$^2$)&$U_{12}$ (\AA$^2$)&$U_{13}$ (\AA$^2$)&
 $U_{23}$ (\AA$^2$)\\
\hline
Pr1 & 0.0153(5) & 0.0088(4) & 0.0012(14) & -0.0021(3) & 0.0015(4) & -0.0026(5)\\
Pr2 & 0.01000(14) & 0.0062(3) & 0.0082(2) & 0.00099(9) & 0.0000 & 0.0000\\
Pr3 & 0.00978(18) & 0.0105(5) & 0.0096(5) & 0.00124(12) & 0.0000 & 0.0000\\
Pr4 & 0.0113(3) & 0.0040(2) & 0.0042(5) & 0.00077(10) & -0.00017(11) & -0.0025(2)\\
Sr1 & 0.0074(2) & 0.0087(4) & 0.0117(9) & 0.0015(3) & -0.0010(3) & -0.0003(4)\\
Sr2 & 0.0099(8) & 0.0011(7) & 0.0072(9) & 0.0014(3) & 0.0000 & 0.0000\\
Sr3 & 0.0055(3) & 0.0040(9) & 0.0067(10) & 0.0018(5) & 0.0000 & 0.0000\\
Sr4 & 0.00903(19) & 0.0041(2) & 0.0059(5) & 0.00071(7) & 0.00007(10) & -0.0027(2)\\
Ca1 & 0.0074(2) & 0.0087(4) & 0.0117(9) & 0.0015(3) & -0.0010(3) & -0.0003(4)\\
Ca2 & 0.0099(8) & 0.0011(7) & 0.0072(9) & 0.0014(3) & 0.0000 & 0.0000\\
Ca3 & 0.0055(3) & 0.0040(9) & 0.0067(10) & 0.0018(5) & 0.0000 & 0.0000\\
Ca4 & 0.00903(19) & 0.0041(2) & 0.0059(5) & 0.00071(7) & 0.00007(10) & -0.0027(2)\\
Mn1 & 0.00487(10) & 0.00486(18) & 0.0066(3) & 0.00036(4) & 0.00001(5) & -0.00008(16)\\
Mn2 & 0.00472(8) & 0.00536(17) & 0.0067(2) & -0.00030(8) & -0.00026(7) & -0.00017(16)\\
O1 & 0.0112(5) & 0.0122(4) & 0.0224(9) & 0.0072(3) & -0.0022(6) & -0.0018(8)\\
O2 & 0.0162(6) & 0.0156(5) & 0.0186(10) & -0.0021(4) & 0.0002(7) & -0.0020(10)\\
O3 & 0.0253(10) & 0.0079(6) & 0.0070(17) & 0.0032(6) & 0.0000 & 0.0000\\
O4 & 0.0255(13) & 0.0164(14) & 0.0036(19) & -0.0043(6) & 0.0000 & 0.0000\\
O5 & 0.0135(6) & 0.0113(4) & 0.0162(8) & 0.0048(4) & 0.0001(4) & -0.0010(6)\\
O6 & 0.0117(7) & 0.0111(5) & 0.0142(8) & -0.0073(3) & 0.0013(4) & -0.0004(5)\\
O7 & 0.0212(9) & 0.0169(11) & 0.009(2) & 0.0001(3) & -0.0001(3) & -0.0071(12)\\
O8 & 0.0239(9) & 0.0099(6) & 0.0014(19) & 0.0009(3) & -0.0005(3) & -0.0009(7)\\
\end{tabular}
\end{ruledtabular}
\end{table*}

\begin{table*} 
\caption{\label{table_10}
The structure parameters of Pr(Sr$_{0.1}$Ca$_{0.9}$)$_2$Mn$_2$O$_7$ in the charge- and orbital-ordered phase (CO2) 
at 295 K. 
The 13242 reflections were observed, and 8080 of them are independent. 
The 144 variables were used for the refinement. 
The space group is $Am2m$ (No.38) of orthorhombic crystal system. 
The latticce parameters are as follows: $a$=10.8026(7) \AA, $b$=5.4719(4) \AA, $c$=19.2090(10) \AA. 
The reliability factors are $R$= 2.95\, $R_{w}$=3.80\, GOF=1.027. 
}
\begin{ruledtabular}
\begin{tabular}{ccccccc}
 &site&$x$&$y$&$z$&$g$&$B$ (\AA$^2$)\\
\hline
Pr1 & 8$f$ & 0.252310(10) & 0.762490(10) & 0.183810(10) & 0.21 & 0.556(17)\\
Pr2 & 4$e$ & 1/2 & 0.246830(10) & 0.185150(10) & 0.21 & 0.528(13)\\
Pr3 & 4$d$ & 0 & 0.258500(10) & 0.183300(10) & 0.21 & 0.502(12)\\
Pr4 & 4$c$ & 0.248560(10) & 0.757300(10) & 0 & 0.58 & 0.598(4)\\
Pr5 & 2$b$ & 1/2 & 0.249200(10) & 0 & 0.58 & 0.668(7)\\
Pr6 & 2$a$ & 0 & 0.262900(10) & 0 & 0.58 & 0.646(7)\\
Sr1 & 8$f$ & 0.249310(10) & 0.763690(10) & 0.183810(10) & 0.08 & 0.688(15)\\
Sr2 & 4$e$ & 1/2 & 0.219210(10) & 0.183490(10) & 0.08 & 0.485(6)\\
Sr3 & 4$d$ & 0 & 0.244500(10) & 0.183700(10) & 0.08 & 0.669(17)\\
Sr4 & 4$c$ & 0.243700(10) & 0.736500(10) & 0 & 0.04 & 0.55(2)\\
Sr5 & 2$b$ & 1/2 & 0.216600(10) & 0 & 0.04 & 0.354(14)\\
Sr6 & 2$a$ & 0 & 0.274300(10) & 0 & 0.04 & 0.40(3)\\
Ca1 & 8$f$ & 0.249310(10) & 0.763690(10) & 0.183810(10) & 0.71 & 0.688(15)\\
Ca2 & 4$e$ & 1/2 & 0.219210(10) & 0.183490(10) & 0.71 & 0.485(6)\\
Ca3 & 4$d$ & 0 & 0.244500(10) & 0.183700(10) & 0.71 & 0.669(17)\\
Ca4 & 4$c$ & 0.243700(10) & 0.736500(10) & 0 & 0.38 & 0.55(2)\\
Ca5 & 2$b$ & 1/2 & 0.216600(10) & 0 & 0.38 & 0.354(14)\\
Ca6 & 2$a$ & 0 & 0.274300(10) & 0 & 0.38 & 0.40(3)\\
Mn1 & 8$f$ & 0.25029(2) & 0.25753(12) & 0.09946(2) & 1 & 0.357(3)\\
Mn2 & 4$e$ & 1/2 & 0.74255(13) & 0.09886(3) & 1 & 0.461(7)\\
Mn3 & 4$d$ & 0 & 0.76417(13) & 0.10010(2) & 1 & 0.326(4)\\
O1 & 8$f$ & 0.1224(2) & 0.0199(3) & 0.60700(9) & 1 & 1.18(2)\\
O2 & 8$f$ & 0.3764(2) & 0.9883(3) & 0.39083(9) & 1 & 1.129(18)\\
O3 & 8$f$ & 0.1245(2) & 0.0116(4) & 0.08692(7) & 1 & 1.007(16)\\
O4 & 8$f$ & 0.3775(2) & 0.9789(3) & 0.91264(7) & 1 & 0.982(19)\\
O5 & 8$f$ & 0.24888(12) & 0.2069(4) & 0.19839(15) & 1 & 1.07(3)\\
O6 & 4$e$ & 1/2 & 0.7902(4) & 0.19831(17) & 1 & 1.11(3)\\
O7 & 4$d$ & 0 & 0.8081(4) & 0.19837(17) & 1 & 1.21(3)\\
O8 & 4$c$ & 0.25303(19) & 0.3097(5) & 0 & 1 & 1.10(4)\\
O9 & 2$b$ & 1/2 & 0.6791(7) & 0 & 1 & 1.34(5)\\
O10 & 2$a$ & 0 & 0.7105(6) & 0 & 1 & 1.02(4)\\
\end{tabular}
\begin{tabular}{ccccccc}
 &$U_{11}$ (\AA$^2$)&$U_{22}$ (\AA$^2$)&$U_{33}$ (\AA$^2$)&$U_{12}$ (\AA$^2$)&$U_{13}$ (\AA$^2$)&
 $U_{23}$ (\AA$^2$)\\
\hline
Pr1 & 0.0084(4) & 0.0064(4) & 0.0063(4) & -0.0006(3) & 0.0022(2) & -0.0009(3)\\
Pr2 & 0.0053(2) & 0.0080(4) & 0.0067(2) & 0.0000 & 0.0000 & -0.00100(19)\\
Pr3 & 0.0089(4) & 0.0049(2) & 0.0053(2) & 0.0000 & 0.0000 & 0.0009(2)\\
Pr4 & 0.00851(15) & 0.00681(9) & 0.00739(11) & -0.00027(9) & 0.0000 & 0.0000\\
Pr5 & 0.01083(16) & 0.0062(2) & 0.00835(17) & 0.0000 & 0.0000 & 0.0000\\
Pr6 & 0.0113(2) & 0.00492(18) & 0.0084(2) & 0.0000 & 0.0000 & 0.0000\\
Sr1 & 0.0134(4) & 0.0068(4) & 0.0060(3) & 0.0000(3) & -0.0017(2) & -0.0003(2)\\
Sr2 & 0.01022(19) & 0.00283(17) & 0.00537(15) & 0.0000 & 0.0000 & -0.00047(12)\\
Sr3 & 0.0072(4) & 0.0127(5) & 0.0055(3) & 0.0000 & 0.0000 & -0.0006(2)\\
Sr4 & 0.0052(4) & 0.0100(6) & 0.0056(4) & -0.0040(4) & 0.0000 & 0.0000\\
Sr5 & --- & --- & --- & --- & --- & ---\\
Sr6 & --- & --- & --- & --- & --- & ---\\
Ca1 & 0.0134(4) & 0.0068(4) & 0.0060(3) & 0.0000(3) & -0.0017(2) & -0.0003(2)\\
Ca2 & 0.01022(19) & 0.00283(17) & 0.00537(15) & 0.0000 & 0.0000 & -0.00047(12)\\
Ca3 & 0.0072(4) & 0.0127(5) & 0.0055(3) & 0.0000 & 0.0000 & -0.0006(2)\\
Ca4 & 0.0052(4) & 0.0100(6) & 0.0056(4) & -0.0040(4) & 0.0000 & 0.0000\\
Ca5 & --- & --- & --- & --- & --- & ---\\
Ca6 & --- & --- & --- & --- & --- & ---\\
Mn1 & 0.00380(9) & 0.00412(9) & 0.00563(10) & -0.00053(3) & 0.00006(4) & 0.00053(10)\\
Mn2 & 0.00391(14) & 0.0081(2) & 0.00546(16) & 0.0000 & 0.0000 & 0.00078(10)\\
Mn3 & 0.00420(10) & 0.00277(11) & 0.00540(11) & 0.0000 & 0.0000 & 0.00018(11)\\
O1 & 0.0110(4) & 0.0148(6) & 0.0189(4) & 0.0045(3) & 0.0003(6) & 0.0024(4)\\
O2 & 0.0113(4) & 0.0123(4) & 0.0193(4) & -0.0069(3) & 0.0008(5) & -0.0030(4)\\
O3 & 0.0104(3) & 0.0132(5) & 0.0146(3) & -0.0067(3) & -0.0003(4) & -0.0020(4)\\
O4 & 0.0118(4) & 0.0118(6) & 0.0137(4) & 0.0052(3) & 0.0003(3) & 0.0003(3)\\
O5 & 0.0196(9) & 0.0153(7) & 0.0056(7) & -0.0007(3) & 0.0005(2) & -0.0029(5)\\
O6 & 0.0220(11) & 0.0142(8) & 0.0059(7) & 0.0000 & 0.0000 & -0.0031(6)\\
O7 & 0.0262(12) & 0.0137(9) & 0.0060(8) & 0.0000 & 0.0000 & -0.0025(5)\\
O8 & 0.0233(13) & 0.0127(9) & 0.0058(8) & -0.0012(4) & 0.0000 & 0.0000\\
O9 & 0.0285(17) & 0.0181(14) & 0.0044(9) & 0.0000 & 0.0000 & 0.0000\\
O10 & 0.0219(14) & 0.0140(10) & 0.0029(7) & 0.0000 & 0.0000 & 0.0000\\
\end{tabular}
\end{ruledtabular}
\end{table*}

\begin{table*}
\caption{\label{table_11}
Distortion of MnO$_6$ octahedra at 405 K,  330 K, and 295 K of Pr(Sr$_{0.1}$Ca$_{0.9}$)$_2$Mn$_2$O$_7$. 
}
\begin{ruledtabular}
\begin{tabular}{ccccccccc}
405 K &$d_x$ (\AA$^2$)&$d_y$ (\AA$^2$)&$d_z$ (\AA$^2$)&$Q_1$&$Q_2$&$Q_3$&$V$&$\theta$\\
\hline
Mn1 & 1.9309(3) & 1.9309(3) & 1.931(2) & -0.043(2) & 0.000(2) & 0.000(2) & 3.72(2) & 0$^{\circ}$\\
\end{tabular}
\begin{tabular}{ccccccccc}
330 K &$d_x$ (\AA$^2$)&$d_y$ (\AA$^2$)&$d_z$ (\AA$^2$)&$Q_1$&$Q_2$&$Q_3$&$V$&$\theta$\\
\hline
Mn1 & 2.007(4) & 1.899(3) & 1.952(4) & -0.006(3) & 0.076(3) & -0.001(3) & 3.53(2) & 91(3)$^{\circ}$\\
Mn2 & 1.898(4) & 1.925(5) & 1.908(4) & -0.080(3) & 0.019(3) & -0.003(3) & 3.92(2) & 10(1)$\times$10$^{\circ}$\\
\end{tabular}
\begin{tabular}{ccccccccc}
295 K &$d_x$ (\AA$^2$)&$d_y$ (\AA$^2$)&$d_z$ (\AA$^2$)&$Q_1$&$Q_2$&$Q_3$&$V$&$\theta$\\
\hline
Mn1 & 2.032(2) & 1.898(2) & 1.926(3) & -0.007(2) & 0.095(2) & -0.032(2) & 3.53(2) & 109(1)$^{\circ}$\\
Mn2 & 1.901(2) & 1.901(2) & 1.929(3) & -0.080(2) & 0.000(2) & 0.023(2) & 3.92(2) & 0(5)$^{\circ}$\\
Mn3 & 1.905(2) & 1.905(2) & 1.924(3) & -0.077(2) & 0.000(2) & 0.016(2) & 3.91(2) & 0(8)$^{\circ}$\\
\end{tabular}
\end{ruledtabular}
\end{table*}


\begin{thebibliography}{200}
\bibitem{footnote2}
The average bond length $\bar{d}$ for Mn$^{3.5+}$ is estimated from the fitting of Valence-distance (Mn-O bond length) curves 
calculated by the bond valence sum of Mn$^{3+}$ and Mn$^{4+}$.
\bibitem{footnote3}
The calculated bond valence sum for oxygen shows a value of $-2\pm0.1$ for all materials, 
indicating that doped number is well included in the calculated nominal valence of Mn ions. 
However, it should be noted that doped holes have strong O 2$p$ character in reality.
\bibitem{footnote1}
Minimal disorder effect is also manifested by the absence of enhancement of 
the atomic displacement factors (see tables in APPENDIX), ensuring  the validity 
of using bond valence sum analysis.
\bibitem{footnote-Vvalue}
Averaged value of $V$ is 3.61 for  Pr$_{0.5}$Ca$_{0.5}$MnO$_3$, and is larger than 
the nominal value of  3.5. Similar tendency is found also for the single layer and the double layer compounds.
This would be due to the compressed (stretched) nature of Mn-O (rare-earth/alkaline-earth-O) bond and  
the resultant charge transfer from Mn site to the rare-earth/alkaline-earth site.
\end{thebibliography}
\end{document}